\documentclass{elsart}

\usepackage{graphicx,epsfig,pstricks,rotating}
\usepackage{amsmath}
\usepackage{amssymb}

\newcommand{\beq}{\begin{equation}}
\newcommand{\eeq}{\end{equation}}
\newcommand{\bdm}{\begin{displaymath}}
\newcommand{\edm}{\end{displaymath}}
\newcommand{\beqa}{\begin{eqnarray}}
\newcommand{\eeqa}{\end{eqnarray}}

\begin{document}

\begin{frontmatter}

\title{Modeling long-range cross-correlations in two-component
ARFIMA and FIARCH processes}

\author[label1,label2]{Boris~Podobnik}
\author[label3]{Davor~Horvatic}
\author[label4]{Alfonso Lam Ng}
\author[label4]{H. Eugene Stanley}
\author[label4,label5,label6]{Plamen~Ch.~Ivanov}

\address[label1]{Department of Physics, Faculty of Civil Engineering, University of Rijeka,
    Rijeka, Croatia}
\address[label2]{Zagreb School of Economics and Management,
 Zagreb, Croatia} 
 
\address[label3]{Department of Physics, Faculty of Science, University of Zagreb, 
Zagreb, Croatia}

\address[label4]{Center for Polymer Studies and Department of
    Physics, Boston University, Boston, MA 02215}

\address[label5]{Division of Sleep Medicine, Brigham and Woman's Hospital, Harvard
Medical School, Boston,   MA 02115}
\address[label6]{Institute of Solid State Physics, Bulgarian Academy of Sciences, 
  Sofia, Bulgaria}

\begin{abstract} 
 We investigate how simultaneously recorded
  long-range power-law correlated 
 multivariate  signals cross-correlate. To this end we introduce a 
 two-component ARFIMA stochastic process and a two-component
  FIARCH process to generate coupled fractal 
  signals with long-range power-law correlations which are at the same time 
  long-range cross-correlated. We study how the degree of cross-correlations 
  between these signals depends on the scaling exponents characterizing the 
  fractal correlations in each signal and on the coupling between the signals. 
   Our findings  have relevance when studying parallel outputs of 
	multiple-component
of physical, physiological and social systems. 
\end{abstract}

\end{frontmatter}


Many empirical data are characterized by 
long-range power-law auto-correlations as well as by 
long-range cross-correlations. 
Such scale-invariant  organization in both auto-correlations and 
cross-correlations can be observed either for the data variables or their
 absolute  values  
\cite{Ying,Crouch,Tauchen,Karpoff2,Gallant,Campbell,Plerou,gopi,Lebaron}.

 Scale-invariant  power-law auto-correlations in stochastic variables 
   can be 
  modeled by  the fractionally autoregressive 
 integrated moving-average process (ARFIMA)
 \cite{frac,Hosk81}:  
    \begin{eqnarray}
    x_t &=&  \sum_{n=1}^{\infty} {a_n(d) }  x_{t-n} 
    +   \epsilon_t,
    \label{arfima1}
  \end{eqnarray}  
  where $d~\in~(-0.5, 0.5)$ is a scaling parameter, 
  $\epsilon_t$ denotes independent and identically
  distributed (i.i.d.) Gaussian variables with  $\langle
  \epsilon_t\rangle =0$ and $\langle\epsilon_t^2 \rangle=1$, 
  $a_n(d)$ are the weights defined by $a_n(d) =
  d ~\Gamma(n-d) / (\Gamma(1-d) \Gamma(n+1))$, where   
$\Gamma$ denotes the
  Gamma function and $n$ is the time scale.   
We denote the auto-correlation function 
for $x_t$ as $A(x_t,x_{t-n}) \equiv A(n)$. For 
$d=0$  the generated variable $x_t$  becomes random.

  To account for power-law  cross-correlations 
  between two variables $x_t$ and
  $y_t$, 
where each variable is itself power-law  auto-correlated, 
 we propose a two-component ARFIMA  stochastic process defined  by two
stochastic variables $x_t$ and $y_t$. 
 Each of these variables at any time depends not
only on its own past values but also on past values of the 
other variable:
\begin{subequations}
\begin{eqnarray}
x_t &=&  [W X_{t}  + (1-W)  Y_{t} ]  +  \epsilon_{t},
\label{Xt}\\
y_t &=& [( 1-W) X_{t}  + W  Y_{t} ]  +  \tilde \epsilon_{t},
\label{Yt}\\
X_{t} &=& \sum_{n=1}^{\infty} 
a_n(d_1)  x_{t-n}, 
\label{Sigx}\\
Y_{t} &=&   \sum_{n=1}^{\infty} 
 a_n(d_2)  y_{t-n},
\label{Sigy}
\end{eqnarray}
\end{subequations}
  where $\epsilon_{t}$ and $ \tilde \epsilon_{t}$  
  denote i.i.d.  Gaussian variables with  $\langle
  \epsilon_t \rangle = \langle
 \tilde  \epsilon_t \rangle =0$ and $\langle  \epsilon_t^2 \rangle =
 \langle  \tilde \epsilon_t^2 \rangle=1$, 
  $a_n(d_1)$ and $a_n(d_2)$ are the 
  weights  defined in Eq.~(1) through the scaling parameters $d_1$ 
  and $d_2$ ($0 \le d_{1,2} < 0.5$),  
  and   $W$ is a free parameter controlling   the coupling strength 
  between  $x_t$ and $y_t$ ($ 0.5 \le W \le 1$).
   We denote the cross-correlation function between
	 $x_t$ and $y_t$ as  $C(x_t,y_{t-n}) \equiv C(n)$. 
 For different values of $W$ a different degree of cross-correlation  
 between the variables $x_t$ and $y_t$ is observed. For example, 
  for  the case  when $W=1$, the process defined in 
   Eqs.~(\ref{Xt})-(\ref{Sigy})
 reduces to two decoupled ARFIMA processes defined in    
Eq.~(\ref{arfima1}). Thus,  when $W=1$ the long-range cross-correlations
  between  $x_t$ and $y_t$ vanish, while  both  $x_t$ and $y_t$ remain 
 long-range power-law auto-correlated.  

\begin{figure}
\centerline{\includegraphics[width=80mm]{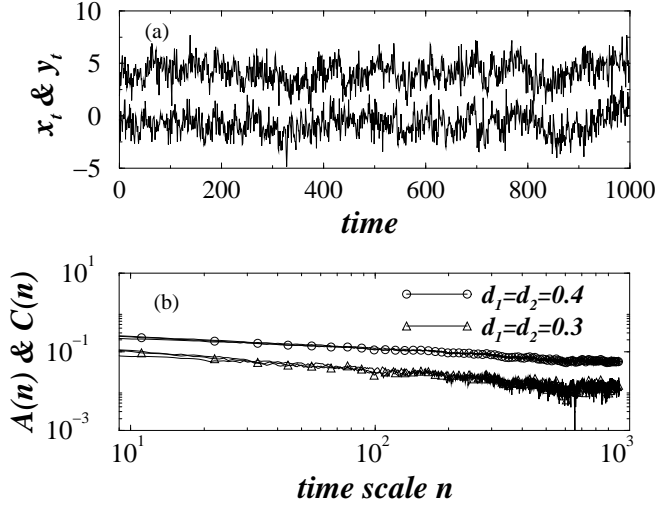}}
\caption{(a) Time series $x_t$ and $y_t$  for the process defined in
 Eqs.(\ref{Xt})-(W{Sigy}) 
where  $W  = 0.8$ and $d_1=d_2=0.4$. The time series $x_t  $
 is vertically shifted for clarity.  Both $x_t$ and $y_t$ exhibit  
apparent comovement, indicating a high degree of cross-correlation.   
 (b)  Log-log plots of the 
auto-correlation  functions $A(n)$ for  $x_t$ and $y_t$,  
 and their cross-correlation function  $C(n)$ for the 
 two-component ARFIMA  process  
with $W  = 0.8$ and $d_1=d_2=0.4$ (top three curves),  and with   
$W  = 0.8$ and  $d_1=d_2=0.3$ (bottom three curves).  
For decreasing values of the scaling parameters $d_1$ and $d_2$  
both the auto-correlations and 
cross-correlations  decrease, leading to smaller values of $A(n)$ and $C(n)$. 
}
\label{fig.1}
\end{figure}

\begin{figure}
\centerline{\includegraphics[width=80mm]{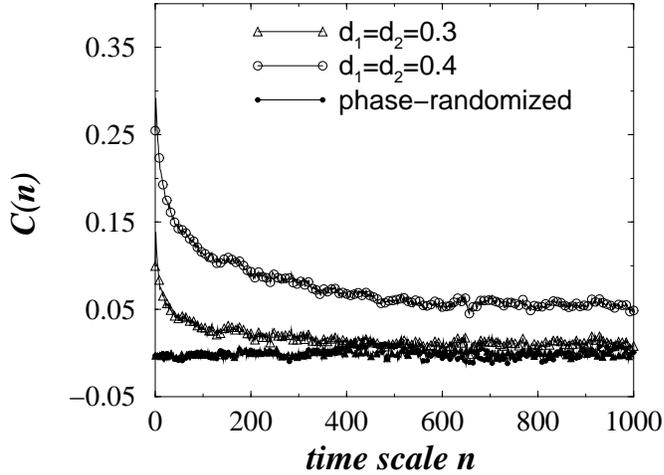}}
\caption{Cross-correlation function $C(n)$  before 
Fourier phase-randomization procedure for 
  the  time series $x_t$ and $y_t$ shown in Fig.~1 (open symbols).
   After Fourier phase randomization 
  of $x_t$ and $y_t$ the cross-correlation
 function virtually  disappears (filled symbols)  for any
  value of $d_1$ and $d_2$. 
}
\label{fig.2}
\end{figure}

In Fig. 1(a) we show  segments  of the time  series   $x_t$ and $y_t$
 generated by the process defined  in  Eq.~(2a)-(2d)
with  parameters   $W= 0.8$ and $d_1=d_2=0.4$.
 Both  variables exhibit a very similar comovement.  
 In Fig. 1(b) we  show the  auto-correlation functions
 $A(n)$ for $x_t$ and $y_t$, 
as well as  the cross-correlation function  $ C(x_{t},y_{t-n}) \equiv C(n) $. 
These three  curves practically  overlap [Fig.~1(b), three top curves].  
We also show the same 
correlation functions for $W= 0. 8$ and $d_1=d_2=0.3$ [Fig.~1(b),  
three bottom 
curves].  Generally, 
when the coupling parameter  $W$ is kept fixed, 
the stochastic process we introduce 
in Eq.~(2) generates stronger cross-correlations 
for larger  values  of the scaling parameters $d_1$ and  $d_2$.

Motivated by the fact that for linear processes the auto-correlation
 function
does not change under randomization of the Fourier phase 
 \cite{Eur,Ashk20}, we next   
test how this phase-randomization procedure 
affects the degree of cross-correlation  between $x_t$ and $y_t$.   
First, we perform a Fourier transform of  
the original time series, e.g. $x_t$, 
preserving the Fourier amplitudes  
but randomizing  the Fourier phases. Then, we perform  
an inverse Fourier transform    
and obtain a surrogate (linearized) time series $\tilde x_t$. 
Applying this phase-randomization procedure to
both time series $x_t$ and $y_t$ generated by the 
two-component ARFIMA process
in Eq. ~(2), we  calculate 
the two auto-correlation functions for
 $\tilde x_t$ and $\tilde y_t$, as well as their  
 cross-correlation function  
$ C(\tilde x_{t},\tilde y_{t-n})$. 
As expected,  the 
 auto-correlation  functions  remain unchanged after Fourier 
 phase randomization, 
 but the 
cross-correlation function $ C(\tilde x_{t},\tilde y_{t-n})$  
completely  vanishes [Fig.~2].

Next, we investigate the case when the scaling parameters  
$d_1$ and $d_2$ are fixed, 
while the coupling parameter $W$  varies.  
In Fig.~3, we  show how the cross-correlation function changes 
  for different values  of  
$W$ and for fixed $d_1 = d_2 = 0.4$. The closer the value of the 
parameter $W$  to 1,  the weaker the cross-correlations  
($W=1 $ corresponds to the case of two 
decoupled  ARFIMA processes).

\begin{figure}
\centerline{\includegraphics[width=80mm]{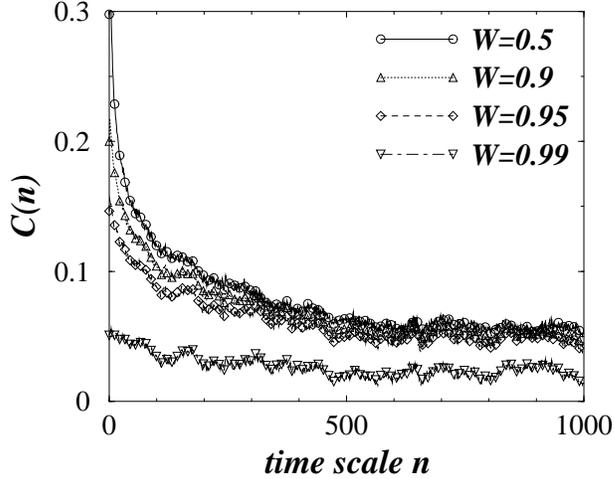}}
\caption{Cross-correlation  function $C(n)$ between time series $x_t$ and $y_t$  
generated by the process in Eqs.(2) for varying values of $W$ and 
 $d_1=d_2=0.4$. 
The cross-correlation function  has highest  values for  $W=0.5$, and 
tends to zero  for $W$ approaching 1.  When $W=1$,  
 $x_t$ and $y_t$ become two decoupled ARFIMA processes.
}
\label{fig.3}
\end{figure}

\begin{figure}
\centerline{\includegraphics[width=80mm]{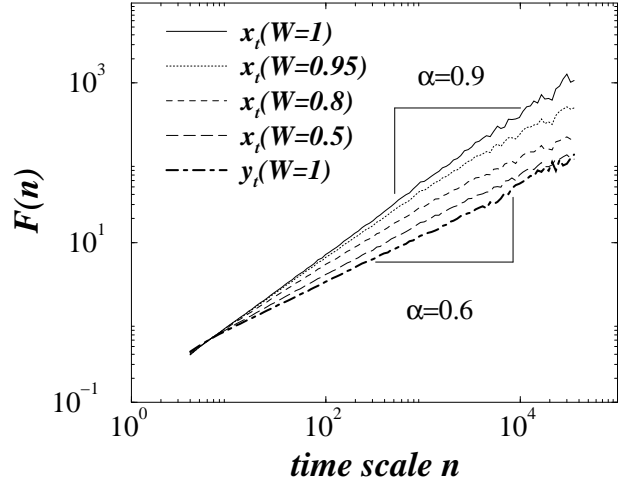}}
\caption{DFA scaling 
curves for  the time series 
$x_t$ and $y_t$  generated by the two-component ARFIMA   
process in Eqs.~(\ref{Xt})-(\ref{Sigy}), 
where $d_1 = 0.4$ and $d_2=0.1$. For $W=1$, $x_t$ and $y_t$  
are decoupled and thus not cross-correlated,  and  $x_t$  behaves as 
the ARFIMA process in Eq.~(\ref{arfima1}) 
defined only by the scaling parameter 
$d_1$,  while 
$y_t$  becomes   a separate  ARFIMA  process defined only by the 
scaling parameter  $d_2$.  
 For $W \neq 1$,  the scaling properties of 
$x_t$ depend on both parameters $d_1$ and $d_2$. 
 When $W=0.5$,   the 
DFA correlation  exponent $\alpha$ 
for $x_t$ becomes equal to the DFA correlation exponent for  
 $y_t$. The DFA exponent for $|y_t|$  does not depend on $W$.
}
\label{fig.4}
\end{figure}

Next we analyze how the  degree of  power-law 
auto-correlations    changes    
when varying the parameters $W$, $d_1$, and $d_2$ in 
Eqs.~(\ref{Xt})-(\ref{Sigy}). 
To quantify the auto-correlations 
we employ the detrended fluctuations  analysis  
(DFA) method. We estimate  the rms fluctuation 
function  $F(n)$  for different time scales $n$ \cite{DFA,DFAa,DFA1,DFA2,DFA3}. 
A power-law dependence  of $F(n)$  on the time scale  $n$ --- 
  $F(n)   \propto n^{\alpha}$, 
 where $\alpha$ is the correlation exponent --- indicates  presence 
 of power law auto-correlations.  
In Fig.~4, we show the DFA scaling curves   obtained for 
$x_t$ and $y_t$ generated by the two-component ARFIMA process
in Eqs.~(\ref{Xt})-(\ref{Sigy}), where $d_1=0.4$ and $d_2 =0.1$, and  the coupling parameter  
$W$  varies. 
 For $W=1$ the processes  $x_t$ and $y_t$ are decoupled and thus
  not cross-correlated. In this case, 
$x_t$ behaves as a power-law auto-correlated
 ARFIMA process controlled by only  the scaling parameter 
$d_1$,  with the DFA correlation exponent equals  
 $\alpha=0.5 +d_1=0.9$. Similarly,  
$y_t$ becomes a separate ARFIMA  process (decoupled from $x_t$) 
which is controlled only by the scaling  parameter  $d_2$, 
 where  $\alpha= 0.5+ d_2=0.6$. We find that 
  with decreasing value of $W$ (from 1 to 0.5), $x_t$  becomes a mixture of 
two ARFIMA processes and the  DFA correlation exponent $\alpha$ 
gradually decreases towards  $\alpha = 0.6$ 
corresponding to the $y_t$ process, controlled 
by parameter $d_2=0.1$. In contrast to $x_t$, for the process $y_t$
the DFA correlation  exponent $\alpha$    
virtually  does not change with  varying  the coupling parameter  $W$.  

We next  consider a separate stochastic process which generates 
simultaneously two time series with 
power-law auto-correlated  absolute 
values of their variables and  long-range  cross-correlations 
between  these absolute values.   
Power-law auto-correlations in the absolute values  of the stochastic variables 
can be modeled by 
  the Fractionally Integrated ARCH (FIARCH) process  
 \cite{Gra95,pod2005}:
\begin{subequations}
\begin{eqnarray}
x_t  &=& \sigma_t \epsilon_t  \\
\label{fiarch1}
\sigma_t &=& 
\sum_{n=1}^{\infty} 
 a_n(d)\frac{|x_{t-n}|}{\mu_x},
\label{fiarch2}
\end{eqnarray}
\end{subequations}
where $\epsilon_t$  denotes an i.i.d. Gaussian variable 
with  $\langle \epsilon_t\rangle =  
0$ and $  \langle\epsilon_t^2 \rangle=1$, and 
 $0<d<1/2$ and $\mu_x = \langle |x_t| \rangle $. 
The sum of the weights $a_n(d)$ satisfies
$ \sum_{n=1}^{\infty} \frac{d~\Gamma(n-d)}
{\Gamma(1-d) \Gamma(n+1)} = 1  $, 
  yielding $ \langle \sigma_t \rangle = 1$.  
While for the time series  $x_t$ generated by Eq.~(1)  
 the autocorrelation function      
 $A(x_t, x_{t-n})$ is zero for all time scales $n$, 
for the absolute values $|x_t|$ the auto-correlation function is 
 $A(|x_t|, |x_{t-n}|) = 
\Gamma(1-d) \Gamma(n+d) /( \Gamma(d) \Gamma(n+1-d)  )$, which for 
  $n>>1$  converges to the power law 
 $ A(n) \sim n^{-1+2d}$.


To account for power-law  cross-correlations between 
the absolute values of two variables, 
where the absolute values of  each variable 
are simultaneously power-law   auto-correlated, 
we have previously introduced \cite{Eur07}  
a two-component FIARCH process    
with  scaling parameters 
$d_1$ and $d_2$:
\begin{subequations}
\begin{eqnarray}
x_t &=&  [W \sigma_{x t}  + (1-W)  \sigma_{y t} ] \epsilon_{t}  
\label{xt}\\
y_t &=& [( 1-W) \sigma_{x t}  + W  \sigma_{y t} ] \tilde \epsilon_{t}
\label{yt}\\
\sigma_{x t} &=&  \sum_{n=1}^{\infty} \frac{d_1~\Gamma(n-d_1)}
{\Gamma(1-d_1) \Gamma(n+1)}  \frac{|x_{t-n}|}{\mu_x} 
\label{sigx}\\
\sigma_{y t} &=&  \sum_{n=1}^{\infty}\frac{d_2~\Gamma(n-d_2)}
{\Gamma(1-d_2) \Gamma(n+1)} \frac{|y_{t-n}|}{\mu_y}.
\label{sigy}
\end{eqnarray}
\end{subequations}
where $\epsilon_{t}$ and $ \tilde \epsilon_{t}$ are 
i.i.d. variables 
with  $\langle \epsilon_t\rangle = \langle \tilde \epsilon_t\rangle = 
0$ and $  \langle \tilde \epsilon_t^2 \rangle =  \langle\epsilon_t^2 \rangle=1$, 
 $W$ is the coupling parameter controlling 
the degree of  cross-correlations, 
and   $\mu_x = \langle |x_t| \rangle $ and $ \mu_y = \langle
|y_t|\rangle $.

Note, that each of the variables is controlled by a 
composite volatility --- e.g.  for $x_t$ the composite volatility
is $W_1 \sigma_{x t} + (1 - W_1) \sigma_{y t}$ [Eq.~(\ref{xt})] ---  
 that is a combination of two FIARCH volatilities  
 $\sigma_{x t}$ and  $\sigma_{x t}$  [Eq.~(\ref{fiarch2})]. 
Stability of the FIARCH process is achieved through the condition 
 $\langle \sigma_t \rangle=1$.   To retain  stability for the 
two-component FIARCH process in Eq.~(4), the average values of   the composite 
volatilities $W \sigma_{x t}  + (1-W)  \sigma_{y t}$ 
and  $( 1-W) \sigma_{x t}  + W  \sigma_{y t}$ 
 in  Eqs.~(\ref{xt})-(\ref{yt}) should be  1.   
For $W=1$ the process in  
Eqs.~(\ref{xt})-(\ref{sigy}) reduces  to two  decoupled  
  FIARCH process as defined in 
Eqs.~(\ref{fiarch1})-(\ref{fiarch2}), and thus
$|x_t|$ and $|y_t|$ are  not  cross-correlated.

In Ref.~\cite{Eur07} we  have analyzed  the 
 cross-correlation functions between $|x_t|$  and $|y_t|$
for the process defined in Eqs.~(\ref{xt})-(\ref{sigy}) 
for  varying  values of the 
parameters $W$, $d_1$, and $d_2$.

\begin{figure}
\centerline{\includegraphics[width=80mm]{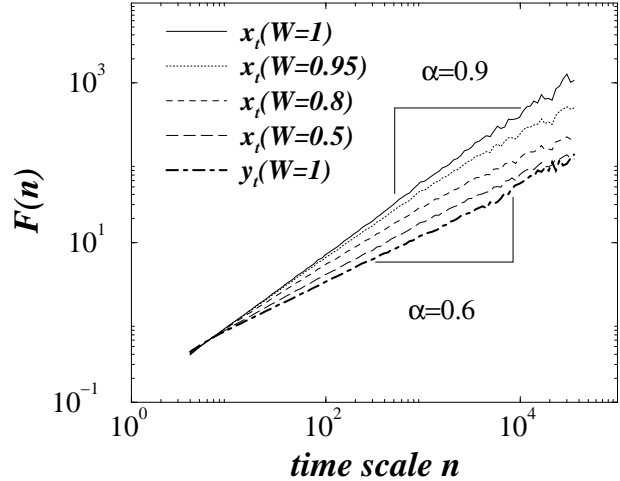}}
\caption{DFA scaling  curves 
of the time series  $|x_t|$ and $|y_t|$  generated by the
two-component FIARCH   
process in Eqs.~(\ref{xt})-(\ref{sigy}), 
where $d_1 = 0.4$ and $d_2=0.1$. For $W=1$, $|x_t|$ and $|y_t|$  
are decoupled and thus are  not cross-correlated. In this case, 
 $x_t$  becomes a separate  FIARCH process as defined in   
Eqs.~(3), and the auto-correlation
 properties of $x_t$ depend only on  the scaling parameter  $d_1$, while 
$y_t$  is another FIARCH process with auto-correlation 
  properties  depending only on  the parameter 
$d_2$. 
 For $W \neq 1$,  the scaling properties of 
$x_t$ depend on both parameters $d_1$ and $d_2$. 
 When $W=0.5$,   the 
DFA correlation  exponent $\alpha$ 
for $|x_t|$ becomes equal to the DFA correlation exponent for  
 $|y_t|$. Note that the 
  DFA exponent for $|y_t|$  does not depend on $W$.
}
\label{fig.5}
\end{figure}

Finally, we analyze how the auto-correlations in the absolute values
  change   
when varying  the parameters $W$, $d_1$, and $d_2$. 
In Fig.~5, we show the DFA  scaling curves   
for $d_1=0.4$ and $d_2 =0.1$, and for varying  
$W$.  For $W=1$,  the  time series
  $x_t$ and $y_t$ are decoupled and so not cross-correlated. 
In this case, $x_t$ is a FIARCH process 
controlled only by the  scaling parameter 
$d_1$, and  exhibits long-range power-law  auto-correlations 
characterized by a 
 DFA correlation exponent   $\alpha=0.5+d_1=0.9$. 
 Similarly,  
$y_t$ is  another FIARCH process controlled only by  $d_2$, and
characterized by
 $\alpha= 0.5+ d_2=0.6$. We find that 
  with decreasing value of $W$ (from 1 to 0), $x_t$  is controlled 
  by  both parameters $d_1$ and $d_2$, 
 and the  DFA exponent $\alpha$ gradually decreases towards 
the value $\alpha = 0.6$. 
 At the same time, the process $y_t$ which is controlled 
only by  the parameter $d_2=0.1$ is also characterized 
by $\alpha = 0.6$, regardless of the values of  $W$.

The presented  modeling approach and findings may have relevance 
when quantifying cross-correlations in simultaneously  recorded multivariate 
time series of fractal nature. This  problem  is pertinent 
to multiple component physical \cite{Gla,God,Camp}, physiological, social  
and financial
 systems.

 We thank the Ministry of Science of Croatia, 
 NIH (Grant HL071972) and NSF for 
 financial support.


\vspace*{-0.3cm}


\begin{thebibliography}{10}


\bibitem{Ying} C. C. Ying,    { Econometrica}  {\bf 34},  (1966) 676.  

\bibitem{Crouch} R. L. Crouch, {  Financial Analystss Journal} {\bf 26}  
(1970) 104.

\bibitem{Tauchen} G. Tauchen  and M. Pitts,  
{ Econometrica}  {\bf 51}, (1983)  485. 


\bibitem{Karpoff2}  J. Karpoff,
{ Journal of  Financial and Quantitative Analysis}  {\bf 22}  (1987) 109.

\bibitem{Gallant} R. Gallant,  P. Rossi, and G. Tauchen, 
 { Review of Financial Studies} {\bf 5}  (1992)  199.
 
\bibitem{Campbell} J. Campbell, A. W.  Lo  and  A. MacKinlay, 
{The Econometrics of Financial Markets}
Princeton NJ: Princeton University Press (1997). 
 
 \bibitem{Plerou} V. Plerou {\it et al.},  
	{Quantitative Finance}  {\bf 1} (2001) 262.  


\bibitem{gopi} P. Gopikrishnan {\it et al.},  
 { Physical Review } E {\bf 62}  (2000)  4493.

\bibitem{Lebaron} B. LeBaron, W. B. Arthur, and  R. Palmer, 
{ Journal of Economic Dynamic \& Control}  {\bf 23},  (1999) 1487.



\bibitem{frac} C. W. J. Granger and R. Joyeux, 
{ J. Time Series   Analysis } {\bf 1},  (1980) 15.
    
\bibitem{Hosk81} J. Hosking, { Biometrika}  {\bf68}, (1981) 165.


\bibitem{pod2005} B. Podobnik  {\it et al.}, 
{  Phys. Rev.} E {\bf 72}, (2005) 026121.
\bibitem{Eur} J. Theiler  {\it et al.},  { Physica} D {\bf 58},  (1992)
 77.

\bibitem{Ashk20} 
 Y. Ashkenazy {\it et al.}, 
{ Physica} A {\bf 323},  (2003) 19.
\bibitem{DFA} 
C.-K. Peng {\it et al.}, Phys. Rev. E {\bf 49}, (1994) 1685.
\bibitem{DFAa} 
 K. Hu  {\it et al.}, 
{ Phys. Rev. E } {\bf 64(1)}  (2001)  011114(19). 
 \bibitem{DFA1}  Z. Chen   {\it et al.}, 
{Phys. Rev. E } {\bf 65(4)}  (2002)  041107(15).  

 \bibitem{DFA2} Z. Chen   {\it et al.}, 
{ Phys. Rev. E}  {\bf 71(1)}  (2005) 011104(11). 
 
 \bibitem{DFA3} 
 L. Xu   {\it et al.}, 
{ Phys. Rev. E}    {\bf 71(5)}   (2005)  051101(14). 

\bibitem{Gra95}  C. W. J. Granger  and Z. Ding  
{ Annales d'Economie et de Statistique}   {\bf 40} (1995) 67.

\bibitem{Eur07} B. Podobnik  {\it et al.}, 
{ Eur. Phys. J.} B {\bf  56}, (2007) 47.

\bibitem{Gla} G. Nugent-Glandorf {\it et al.}, 
{ Phys. Rev. Lett.}  {\bf 87(19)}  (2001)  193002. 

\bibitem{God} O. A. Godin,  
{ Phys. Rev. Lett.}  {\bf 97}  (2006)  054301.

 
\bibitem{Camp} M. Campilo and A. Paul, Science   {\bf 299}  (2003)  547.







\end{thebibliography}
\end{document}